\documentclass[showpacs,pra,aps]{revtex4}
\usepackage{graphicx}
\def\be{\begin{equation}}
\def\ee{\end{equation}}
\def\bea{\begin{eqnarray}}
\def\eea{\end{eqnarray}}
\def\bfr{{\bf r}}

\def\bfs{{\bf s}}
\begin{document}
\title{Exact first-order density matrix 
for a $d$-dimensional harmonically confined Fermi gas at finite temperature}
\author{Brandon P. van Zyl}
\affiliation{Department of Physics and Astronomy,
McMaster University, Hamilton,
Ontario, Canada, L8S~4M1}

\begin{abstract}
We present an exact closed form expression for the {\em finite temperature}
first-order density matrix of a harmonically trapped ideal 
Fermi gas in any dimension.  
This constitutes a much sought after generalization of the recent
results in the literature, where exact expressions have been limited to
quantities derived from the {\em diagonal} first-order density matrix.
We compare our exact results with the Thomas-Fermi approximation (TFA)
and demonstrate numerically that the TFA provides an excellent
description of the first-order density matrix in the large-$N$
limit.
As an interesting application, we derive a closed form 
expression for the finite temperature Hartree-Fock exchange energy of a
two-dimensional parabolically confined quantum dot.  We 
numerically test this exact result
against the 2D TF exchange functional, and comment on the applicability
of the local-density approximation (LDA) to the exchange energy of an inhomogeneous
2D Fermi gas.
\end{abstract}

\maketitle
%%%%%%%%%%%%%%%%%%%%%%%%%%%%%%%%%%%%%%%%%%%%%%
\section{Introduction}
%%%%%%%%%%%%%%%%%%%%%%%%%%%%%%%%%%%%%%%%%%%%%%
The recent technical advances made by DeMarco and Jin~\cite{jin} in the area of
trapped, ultra-cold Fermi gases, has led to the experimental
realization of what is close to being an ideal, non-interacting, many-body system of
harmonically confined fermions.  Using current state-of-the art magneto-optical traps,
it is also now possible to ``tune'' the dimensionality of the gas from three-dimensions (3D)
to quasi-2D or quasi-1D.
Such a model system is of great interest to physicists,
as it provides an opportunity to study the role of dimensionality, and 
the quantum statistical properties of a many-body system, exactly.  
As a result, the last few years have seen
a renewed interest in the theoretical description of harmonically trapped ideal Fermi gases 
at both zero~\cite{vignolo1,gleisberg,brack2,minguzzi1,minguzzi2,vignolo2,howard1,murthy}
and finite temperatures~\cite{akdeniz,brack1}.
The primary focus of these studies has been on examining (analytically and numerically)
the expressions for the {\em local} thermodynamic properties of the gas, e.g., the 
single-particle and kinetic energy densities.
These quantities are, of course, of great importance in the density-functional theory (DFT)
of inhomogeneous Fermi systems, whereby one can by-pass the numerically expensive one-particle
Schr\"odinger equations.  

However, the more fundamental quantity, from which the single-particle and kinetic energy
densities are both derived, is the first-order density matrix $\rho_1(\bfr_1,\bfr_2)$~\cite{parr}.  
Unfortunately, to date,
there are relatively few examples in which a closed analytical form for $\rho_1(\bfr_1,\bfr_2)$
can be written.  One of the earliest examples dates back more than 60 years 
to Husimi~\cite{husimi}, in which
the zero temperature first-order density matrix of a 1D harmonic oscillator
was derived.  Other examples that we are aware of are the so-called Bardeen
model~\cite{bardeen}, corresponding to a planar metal surface, and the work of Bhaduri and 
Sprung~\cite{bhaduri_sprung} dealing with 
a 3D oscillator with a smeared occupancy.  
More recently, Howard {\em et.~al}~\cite{howard} have evaluated the {\em zero temperature}
first-order density matrix of the $d$-dimensional harmonic oscillator for an arbitrary number of
closed shells.  Unfortunately, their form for $\rho_1(\bfr_1,\bfr_2)$ is somewhat impractical in
that it is given in terms of 
multidimensional integrals.   While these integrals are numerically easy to evaluate, they
are not very useful
for further analytical analysis (e.g, examining the asymptotic behaviour of 
$\rho(\bfr)$ as $N\rightarrow \infty$~\cite{brack1,murthy}). 

One of the central theoretical reasons for
pursuing a closed form expression for $\rho_1(\bfr_1,\bfr_2)$ is that its off-diagonal elements
determine the exchange integrals of two-body operators, and hence, the nonlocal properties of the system.
From an experimental point of view, a closed form expression for $\rho_1(\bfr_1,\bfr_2)$ is
desirable because, e.g., the momentum density $n({\bf p})$ (which is just the Fourier transform of the
first-order density matrix), is experimentally accessible by measuring the line-shape in
Compton scattering.  Thus, in the case of a weakly-interacting harmonically confined Fermi gas, an exact
knowledge of $\rho_1(\bfr_1,\bfr_2)$ can serve as a benchmark
from which the effects of interparticle interactions may be extracted.

The primary goal of this paper then, is to present an {\em exact} closed form analytical expression
for the finite temperature first-order density matrix of a harmonically confined ideal Fermi 
in any dimension.  We organize our paper as follows.
In Sec.~II, we briefly review some of the basic
definitions given in~\cite{brack2,brack1}, and then proceed to derive a closed form expression for the
finite temperature first-order density matrix in arbitrary dimensions.  Following this calculation,
we compare our exact $\rho_1(\bfr_1,\bfr_2)$ with the TFA in 2D and
discuss the applicability of the LDA for describing the nonlocal
properties of the 2D trapped gas.  Then, in 
Sec.~IV, we apply our results to construct a closed form expression for
the finite temperature Hartree-Fock (HF) exchange energy density of a parabolically 
confined 2D quantum dot.  
We numerically investigate this exact exchange energy density and comment on its applicability
in the context of the LDA.
Finally, in Sec.~V, we present a summary of our results and briefly highlight
some interesting avenues for further investigation.
%%%%%%%%%%%%%%%%%%%%%%%%%%%%%%%%%%%%%%%%%%%%
%%%%%%%%%%%%%%%%%%%%%%%%%%%%%
\section{First-order density matrix in $d$-dimensions}
\label{fermidensity}
In keeping with our our earlier work~\cite{brack1,brack2}, we begin by considering
a system of noninteracting fermions at zero temperature 
described by the time-independent Schr\"odinger equation
\be
{\hat H} \phi_i(\bfr) = [\hat{T} + V(\bfr)]\phi_i(\bfr) = \varepsilon_i
\phi_i(\bfr)~,
\ee
where $V(\bfr)$ is a one-body potential to be specified later (all 
$\epsilon_i$'s are taken to be positive).   
The (spinless) first-order density matrix can be obtained by an inverse 
Laplace transform
of the zero-temperature Bloch density matrix, $C_0(\bfr_1,\bfr_2)$:
\be
\rho_1(\bfr_1,\bfr_2) = 2 \sum_{\varepsilon_i < E_F}\phi_i^{\star}(\bfr_2)
\phi_i(\bfr_1) \Theta(E_F) = {\cal L}_{E_F}^{-1}\left[\frac{2}{\beta}
C_0(\bfr_1,\bfr_2;\beta)
\right]~,
\label{onesided}
\ee
where 
\be
C_0(\bfr_1,\bfr_2;\beta) = \sum_{{\rm all~} i}\phi_i^{\star}(\bfr_2)\phi_i(\bfr_1)
\exp(-\beta\varepsilon_i)~,
\label{bloch0}
\ee
and $E_F$ is the Fermi energy; the factor of two accounts for spin.
We have 
put in the unit step-function $\Theta(E_F)$ in Eq.~(\ref{onesided}) so that 
the Laplace transform with respect to $E_F$ may formally be taken to be 
two-sided~\cite{pol}.
Note that in quantum statistical mechanics, $\beta$ is usually identified with the
inverse temperature, $\beta = 1/k_B T$.
However, in our present context, $\beta$ is to be interpreted as mathematical 
variable which in general is taken to be complex, and {\it not} the inverse 
temperature $1/k_BT$.

At finite temperature, the first-order density matrix is 
obtained from
the Bloch density matrix by using the relation~\cite{brack}
\be
\rho_1(\bfr_1,\bfr_2;T) = {\cal L}^{-1}_{\mu}\left[ \frac{2}{\beta}
C_T(\bfr_1,\bfr_2;\beta)\right]~,
\label{rhoT}
\ee
where
\be
C_T(\bfr_1,\bfr_2;\beta) = C_0(\bfr_1,\bfr_2;\beta) 
\frac{\pi\beta T}{\sin(\pi\beta T)}~,
\label{blochT}
\ee
is the finite temperature Bloch density matrix, and $\mu$ is the 
chemical potential.  
In Eq.~(4), the Laplace transform with respect to 
$\mu$ is two-sided, so that $\mu$ is allowed to go negative.  
Specializing now to the case of an isotropic harmonic oscillator in
$d$ dimensions, viz.,
\be
V(\bfr) = \frac{1}{2}m\omega^2 r^2,~~~~~~r = \sqrt{x_1^2 + x_2^2 + \cdot
\cdot \cdot+x_d^2}~,
\ee
we have for the zero-temperature Bloch density matrix~\cite{wilson}

\be
C_0^{(d)}(\bfr_1,\bfr_2;\beta) = 
C^{(d)}_0\left(\bfr+\frac{\bfs}{2},\bfr-\frac{\bfs}{2};\beta\right) = 
\left(\frac{1}{2\pi}\right)^{d/2}\frac{1}{\sinh^{d/2}(\beta)}
\exp\left\{-\left[r^2\tanh(\beta/2) + (s^2/4)\coth(\beta/2)\right]\right\}~.
\label{bloch0h}
\ee
In the above expression (and what follows), all lengths and energies have 
been scaled by 
$l_{osc} = \sqrt{\hbar/m\omega}$ and $\hbar\omega$, respectively, and we
have introduced the center-of-mass and relative coordinates:
\be
{\bf r} = \frac{1}{2}(\bfr_1 + \bfr_2),~~~~~{\bf s} = \bfr_1 - \bfr_2~.
\ee

The finite temperature density matrix can, in principle, be obtained by
performing the inverse Laplace transform
given by Eq.~(\ref{rhoT}) with Eq.~(7).  However, rather than following this
direct approach (which is a very difficult task), we first
consider the following identities
\bea
\exp\{-x\tanh(\beta/2)\} &=& \sum_{n=0}^{\infty}(-1)^n L_n(2x)e^{-x}
\{e^{-n\beta} + e^{-(n+1)\beta}\}\nonumber\\
\exp\{-y\coth(\beta/2)\} &=& \sum_{k=0}^{\infty} L_k(2y)e^{-y}
\{e^{-k\beta} - e^{-(k+1)\beta}\}~.
\label{tanh}
\eea
Identifying $x=r^2$ and $y=s^2/4$, and using (9) in Eq. (7), the Bloch
density matrix now reads
\bea
C_0^{(d)}(x,y;\beta) &=& \left(\frac{1}{2\pi}\right)^{d/2}
\frac{1}{\sinh^{d/2}(\beta)}
\sum_{k=0}^{\infty}\sum_{n=0}^{\infty}(-1)^n L_n(2x)L_k(2y)e^{-(x+y)}
\{(e^{-n\beta} + e^{-(n+1)\beta})(e^{-k\beta} - e^{-(k+1)\beta})\}\nonumber \\
&=& \left(\frac{1}{2\pi}\right)^{d/2}
\frac{1}{\sinh^{d/2}(\beta)}
\sum_{k=0}^{\infty}\sum_{n=0}^{\infty}(-1)^n L_n(2x)L_k(2y)e^{-(x+y)}
\{e^{-(n+k)\beta} - e^{-(n+k+2)\beta}\}~.
\label{uni0}
\eea
Substituting Eq.~(\ref{uni0}) into Eq.~(\ref{blochT}) and performing the 
inverse Laplace transform, Eq.~(\ref{rhoT}),
leads to the finite temperature first-order density matrix
$\rho_1(\bfr_1,\bfr_2;T)$ in any dimension.
For the sake of clarity, we will now proceed to give an explicit 
calculation for the simplest case of 2D, 
followed by a statement of the general result in arbitrary dimensions.
%%%%%%%%%%%%%%%%%%%%%%%%%%%%%%%%%%%%
\subsection{Two dimensions}
We begin by noting the
following important exact inverse Laplace transforms (all two-sided):
\be
{\cal L}^{-1}_{\eta}\left[ \frac{e^{-n\beta}}{\sinh(\beta)}\right]
= 2 \sum_{k=0}^{\infty} \delta (\eta - (2k+1) - n)~\Theta(\eta),
\label{twosided}
\ee
\be {\cal L}^{-1}_{\mu} \left[\frac{\pi T}{\sin(\pi\beta T)}\right] =
\frac{1}{\left[\exp(-\frac{\mu}{T})+1\right]}~.
\label{lala}
\ee
Putting $d=2$ in Eq.~(\ref{uni0}) and using Eqs.~(\ref{rhoT}), (\ref{blochT}), 
the finite temperature first-order density matrix
is given by
\be
\rho_1^{(2)}(x,y;T) = \left(\frac{1}{\pi}\right) \sum_{k=0}^{\infty}
\sum_{n=0}^{\infty}
(-1)^n L_n(2x)L_k(2y)e^{-(x+y)}\\
\times {\cal L}_{\mu}^{-1}\left[
\left(\frac{e^{-(n+k)\beta} - e^{-(n+k+2)\beta}}{\sinh(\beta)}\right)
\frac{\pi T}{\sin(\pi\beta T)}\right]
\ee
Applying the convolution theorem for Laplace transforms 
and making use of Eqs.~(\ref{twosided}), (\ref{lala}), we immediately obtain
\bea
\rho_1^{(2)}(x,y;T) &=& \left(\frac{2}{\pi}\right) 
\sum_{k=0}^{\infty}\sum_{n=0}^{\infty}
(-1)^n L_n(2x)L_k(2y)e^{-(x+y)}\times 
\sum_{l=0}^{\infty}\left\{ \int_{-\infty}^{\infty} d\tau 
\delta(\tau - (2l+1)-(n+k))\frac{1}{\left[\exp(\frac{\tau-\mu}{T})+1\right]}\right.
\nonumber \\&-& \left. 
\int_{-\infty}^{\infty} d\tau
\delta(\tau - (2l+3)-(n+k))\frac{1}{\left[\exp(\frac{\tau-\mu}{T})+1\right]}
\right\}\nonumber \\
&=& \left(\frac{2}{\pi}\right) \sum_{k=0}^\infty\sum_{n=0}^{\infty}
(-1)^n L_n(2x)L_k(2y)e^{-(x+y)}\times
\sum_{l=0}^{\infty}\left\{\frac{1}{\left[\exp(\frac{2l+n+k+1-\mu}{T})+1\right]}
-\frac{1}{\left[\exp(\frac{2l+n+k+3-\mu}{T})+1\right]}
\right\}\nonumber \\
&=&
\left(\frac{2}{\pi}\right) \sum_{k=0}^\infty\sum_{n=0}^{\infty}
(-1)^n L_n(2x)L_k(2y)e^{-(x+y)}\times
\frac{1}{\left[\exp(\frac{\varepsilon_n+k-\mu}{T})+1\right]}
\nonumber \\
&=&
\left(\frac{2}{\pi}\right) \sum_{k=0}^\infty\sum_{n=0}^{\infty}
F^{(2)}_{n,k}(\mu)
(-1)^n L_n(2x)L_k(2y)e^{-(x+y)}~,
\label{lily}
\eea
where the function $F^{(2)}_{n,k}(\mu)$ is defined as
\be
F^{(2)}_{n,k}(\mu) \equiv
\frac{1}{\left[\exp(\frac{\varepsilon_n+k-\mu}{T})+1\right]}~,
\label{lucky}
\ee
and $\varepsilon_n = (n+1)$ is the noninteracting energy spectrum 
(in scaled units) of the 2D harmonic
oscillator potential.  Putting back our original variables, we finally arrive
at the following simple expression
\bea
\rho_1^{(2)}(\bfr_1,\bfr_2;T) &=& \left(\frac{2}{\pi}\right) 
\sum_{k=0}^\infty\sum_{n=0}^{\infty} 
F^{(2)}_{n,k}(\mu)
(-1)^n L_n\left(\frac{|\bfr_1+\bfr_2|^2}{2}\right)
L_k\left(\frac{|\bfr_1-\bfr_2|^2}{2}\right)\nonumber \\
&\times&\exp\left(-\left|\frac{\bfr_1+\bfr_2}{2}\right|^2 - \frac{|\bfr_1-\bfr_2|^2}{4}
\right)~.
\label{fine}
\eea
In terms of the center-of-mass and relative coordinates the density matrix
is given by
\bea
\rho_1^{(2)}\left(\bfr+\frac{\bfs}{2},\bfr-\frac{\bfs}{2};T\right)&=& \left(\frac{2}{\pi}\right) \sum_{k=0}^\infty\sum_{n=0}^{\infty}
F^{(2)}_{n,k}(\mu)
(-1)^n L_n\left(2r^2\right)
L_k\left(\frac{s^2}{2}\right)
\exp\left(-\left(r^2 + \frac{s^2}{4}\right)\right)
\eea
Therefore, the density-matrix only depends on the modulus of
$\bfr$ and $\bfs$, and there 
is a clean separation of the variables.
Notice that Eq.~(\ref{fine}) has only two sums owing to the cancellation
of all but the $l=0$ term in the $l$-sum of Eq.~(14).  In addition, it is readily seen that
by setting $\bfr = \bfr_1 = \bfr_2$, we immediately obtain the finite temperature
single-particle density
\be
\rho_1^{(2)}(\bfr_1,\bfr_1) \equiv 
\rho^{(2)}(\bfr;T) = \left(\frac{2}{\pi}\right)
\sum_{n=0}^{\infty}
F^{(2)}_{n}(\mu)(-1)^n L_n(2r^2)
e^{-r^2}~,
\ee
with
\bea
F^{(2)}_n(\mu) &\equiv& \sum_{k=0}^{\infty} F^{(2)}_{n,k}(\mu)\nonumber \\
&=&
\sum_{k=0}^\infty\frac{1}{\left[\exp(\frac{\varepsilon_n+k-\mu}{T})+1\right]}~.
\eea
Equation (18) is of course identical to the result obtained in 
Ref.~\cite{brack1} where only the diagonal part of the first-order density
matrix was investigated. 
Furthermore, the exact 2D zero temperature density matrix can
be obtained from Eq.~(\ref{fine}) by taking the $T\rightarrow 0$ limit, and when
filling $M+1$ shells, reduces to (with all dimensional constants recovered)
\bea
\rho^{(2)}_1(\bfr_1,\bfr_2) &=& \frac{2m\omega}{\pi\hbar }\sum_{n=0}^M
(-1)^n L_n\left(\frac{m\omega}{2\hbar}|\bfr_1+\bfr_2|^2\right)
L_{M-n}^{1}\left(\frac{m\omega}{2\hbar}|\bfr_1-\bfr_2|^2\right)
\exp\left(-\frac{m\omega}{2\hbar}(r_1^2 + r_2^2)\right)\nonumber \\
&=& \frac{2m\omega}{\pi\hbar }\sum_{n=0}^M
(-1)^n L_n\left(\frac{2 m\omega}{\hbar}r^2\right)
L_{M-n}^{1}\left(\frac{m\omega}{2\hbar}s^2\right)
\exp\left(-\frac{m\omega}{\hbar}\left(r^2 + \frac{s^2}{4}\right)\right)
\eea
where $L_{M-n}^1(x)$ is an associated Laguerre polynomial~\cite{gr} [see also Eq.~(34)]
which has the property
that $L_{M-n}^1(0) = M-n+1$.  It is straightforward
to show [with the aid of Eq.~(35)] that Eq.~(20) is indeed idempotent, viz.,
\be
N = \int\int \rho_1^{(2)}(\bfr_1,\bfr_2)\rho_1^{(2)}(\bfr_2,\bfr_1)~d\bfr_1 d\bfr_2~.
\ee
%%%%%%%%%%%%%%%%%%%%%%%%%%%%%%%%%%%%%%
\subsection{Arbitrary dimensions}
The derivation of the first-order density matrix in arbitrary dimensions
closely parallels that of the 2D case.  The few extra steps required
to obtain $\rho_1^{(d)}(\bfr_1,\bfr_2;T)$ 
are clearly laid out in Ref.~\cite{brack1}, and so here, we simply state the final result:
\be 
\rho_1^{(d)}(\bfr_1,\bfr_2;T) = 2\left(\frac{1}{\pi}\right)^{d/2}
\sum_{k=0}^\infty\sum_{n=0}^{\infty}F_{n,k}^{(d)}(\mu) 
(-1)^n L_n\left(\frac{|\bfr_1+\bfr_2|^2}{2}\right)
L_k\left(\frac{|\bfr_1-\bfr_2|^2}{2}\right)
\exp\left(-\frac{1}{2}(r_1^2 + r_2^2)\right)~,
\label{high}
\ee
where
\be
F_{n,k}^{(d)}(\mu) \equiv
\left( \frac{1}{\exp[(\varepsilon^{(d)}_n+k-\mu)/T]+1} +
\sum_{m=1}^{\infty}\frac{g_m^{(d)}}{\exp[(\varepsilon^{(d)}_n+k+2m -\mu)/T]+1}
\right)~,
\ee
and $\varepsilon^{(d)}_n = n + d/2$.
The expansion coefficients can be given in the compact form
\bea
g_m^{(d)} &=& \frac{1}{m!} \frac{\Gamma(d/2+m-1)}{\Gamma(d/2-1)}~.
\eea
In particular, we observe that $g_m^{(2)} = 0$ for all $m$.
We have verified the correctness of Eq.~(22) at $T=0$ by checking that it is an exact
solution of the partial differential equation (valid in any dimension)
\be
\frac{1}{\xi\eta}\frac{\partial^2\rho_1(\xi,\eta)}{\partial\xi\partial\eta} =
\frac{4 m^2 \omega^2}{\hbar^2} \rho_1(\xi,\eta)~,
\ee
where following the notation of \cite{howard}, we have defined
$\xi = |\bfr_1+\bfr_2|/2$ and $\eta = |\bfr_1-\bfr_2|/2$ and
\bea
\rho^{(d)}_1(\xi,\eta) = 2\left(\frac{1}{\pi}\right)^{d/2}
\sum_{n=0}^M(-1)^n L_n(2\xi^2)\left[L_{M-n}^1(2\eta^2) + \sum_{m=1}^{(M-n)/2}
g_m^{(d)}L_{M-n-2m}^1(2\eta^2)\right]e^{-(\xi^2+\eta^2)}~.\nonumber
\eea

Equation (22) also provides us with a
direct way to calculate the finite temperature kinetic energy density
in any dimension, viz.,
\be
\xi^{(d)}(\bfr;T) = -\frac{1}{2}\nabla_s^2\rho_1^{(d)}
\left(\bfr+\frac{\bfs}{2},\bfr - \frac{\bfs}{2};T\right)_{s = 0}~.
\ee
This is an easier route compared with our
previous evaluation of $\xi^{(d)}(\bfr;T)$,
which required an additional inverse Laplace transform of the Bloch 
density matrix at finite temperature (see Eq.~(33) of Ref.~\cite{brack1}).
It should be noted that when making contact with experimental studies
on fully spin-polarized trapped Fermi atoms, it is appropriate to
focus on singly filled levels, which implies that the factor of two
appearing in Eq.~(22) should be dropped.

We also briefly mention that an analogous expression for 
$\rho_1^{(d)}(\bfr_1,\bfr_2;T)$
for the harmonically trapped (spinless) 
Bose gas in any dimension can also be obtained
by simply removing the factor of two in front of Eq. (22) and changing the $+1$
to $-1$ in the denominator of Eq. (23).  While this resulting novel 
expression for the Bose gas is very different from what is found in
the literature, it can be shown to be entirely equivalent to the 
more commonly used form~\cite{brack1}
\bea
\rho^{(d)}_{1}(\bfr_1,\bfr_2;T) &=& \sum_{j=1}^{\infty} \frac{e^{j\mu/T}}
{[\pi(1-e^{-2j/T})]^{d/2}} \times \nonumber \\
&&\exp \left(-\frac{|\bfr_1 +\bfr_2|^2}{4}\tanh(j/2T)-\frac{|\bfr_1-\bfr_2|^2}{4}\
\coth(j/2T)\right)~.
\eea
%%%%%%%%%%%%%%%%%%%%%%%%%%%%%%%%%%%%%%%%%%%%%%%%%%%%%%%%%%%%%%%%%%%%%%%
\subsection{Second-order density matrix}
In the present context of non-interacting fermions, the
second-order density matrix, $\rho^{(d)}_2(\bfr_1\bfr_2,\bfr_1'\bfr_2')$ can immediately be 
obtained from our knowledge of $\rho_1^{(d)}(\bfr_1,\bfr_2)$
(this is in the spirit of the HF approximation)~\cite{parr}.   An important
quantity derived from $\rho_2$ is its diagonal element, which corresponds to the pair density, 
and is given by:
%\be
%\rho^{(d)}_2(\bfr_1\bfr_2,\bfr_1'\bfr_2') = \frac{1}{2}\left\{
%\rho_1^{(d)}(\bfr_1,\bfr_1')\rho_1^{(d)}(\bfr_2,\bfr_2') - 
%\rho_1^{(d)}(\bfr_1,\bfr_2')\rho_1^{(d)}(\bfr_1',\bfr_2)\right\}~.
%\ee
%The diagonal part of Eq.~(27) corresponds to the pair density, and
%reads
\be
\rho^{(d)}_2(\bfr_1\bfr_2,\bfr_1\bfr_2) = \frac{1}{2}\left\{
\rho^{(d)}(\bfr_1)\rho^{(d)}(\bfr_2) - 
\frac{1}{2}|\rho_1^{(d)}(\bfr_1,\bfr_2)|^2\right\}
~~~~~~{\rm [closed~shell]}~.
\ee
For example, in the HF approximation, the 
electron-electron potential energy is given by
\be
V^{\rm HF}_{ee} = \int\int\frac{\rho^{(d)}_2(\bfr_1\bfr_2,\bfr_1\bfr_2)}
{|\bfr_1 - \bfr_2|}d\bfr_1 d\bfr_2~,
\ee
so that the the first term in (28) is responsible for the classical Coulomb
repulsion and the second term yields the quantum-statistical exchange energy.
%%%%%%%%%%%%%%%%%%%%%%%%%%%%%%%%%%%%%%%%%%%%%%%%%%%%%%%%%%%%%%%%%%%%%%%%
%%%%%%%%%%%%%%%%%%%%%%%%%%%%%%%%%%%%%%%%%%%%%%%%%%%%%%%%%%%%%%%%%%%%%%%%
%\section{Momentum density and the Compton line shape}
%In order to attempt to make some contact with experiment, we now proceed
%to discuss the momentum density $n(\bfr p)$.
%%%%%%%%%%%%%%%%%%%%%%%%%%%%%%%%%%%%%%%%%%%%%%%%%%%%%%%%%%%%%%%%%%%%%%%%
%%%%%%%%%%%%%%%%%%%%%%%%%%%%%%%%%%%%%%%%%%%%%%%%%%%%%%%%%%%%%%%%%%%%%%%%
\section{Comparison with the Thomas-Fermi approximation}
In this section, we compare our exact expression for the
first-order density matrix with that of the 
TFA.  Since the largest deviations between 
the exact and TF results are known to be at low
temperatures (and small particle numbers, especially in low-dimensional 
systems), we will restrict our comparisons to 
zero temperature, and simply state here that the temperature dependence
of the exact $\rho_1(\bfr_1,\bfr_2;T)$ is readily studied, should it be
desirable.
Again, for the sake of simplicity, we will focus explicitly on the 2D 
case, although the extension of the analysis 
to other dimensions is straightforward.

In the TFA, the 2D zero temperature Bloch density matrix
is given by~\cite{brack}
\be
C_{TF}(\bfr_1,\bfr_2;\beta) = \left(\frac{1}{2\pi\beta}\right)
e^{-\beta V[(\bfr_1+\bfr_2)/2]}e^{-|\bfr_1-\bfr_2|^2/2\beta}~.
\ee
\begin{figure}
\resizebox{5in}{6in}{\includegraphics{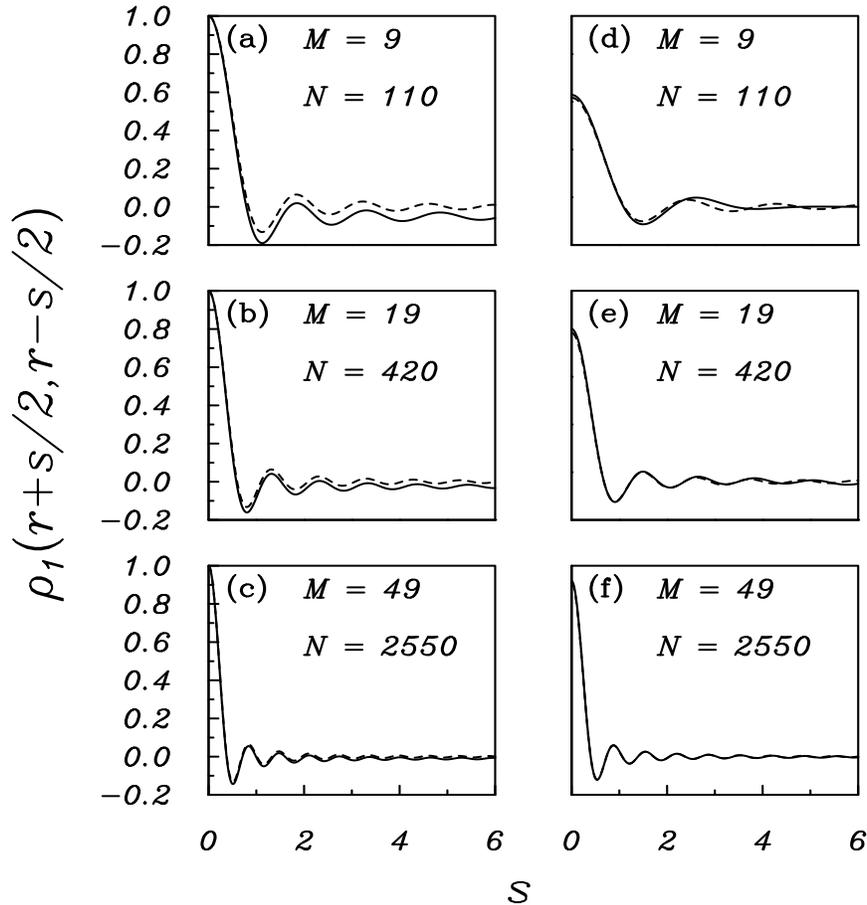}}
\caption{Comparison of the exact zero temperature density matrix
[Eq.~(20) (solid curves)] with the TFA [Eq.~(31) (dashed curves)] 
for various particle numbers.  Panels (a)-(c) correspond to fixing the
center-of-mass coordinate to $\bfr=0$,
and panels (d)-(f) have $\bfr = 3$.  In all cases, $\rho_1$
is normalized to the central density in the trap, $\rho(0)$.}
\label{fig1}
\end{figure}
\begin{figure}
\resizebox{5in}{6in}{\includegraphics{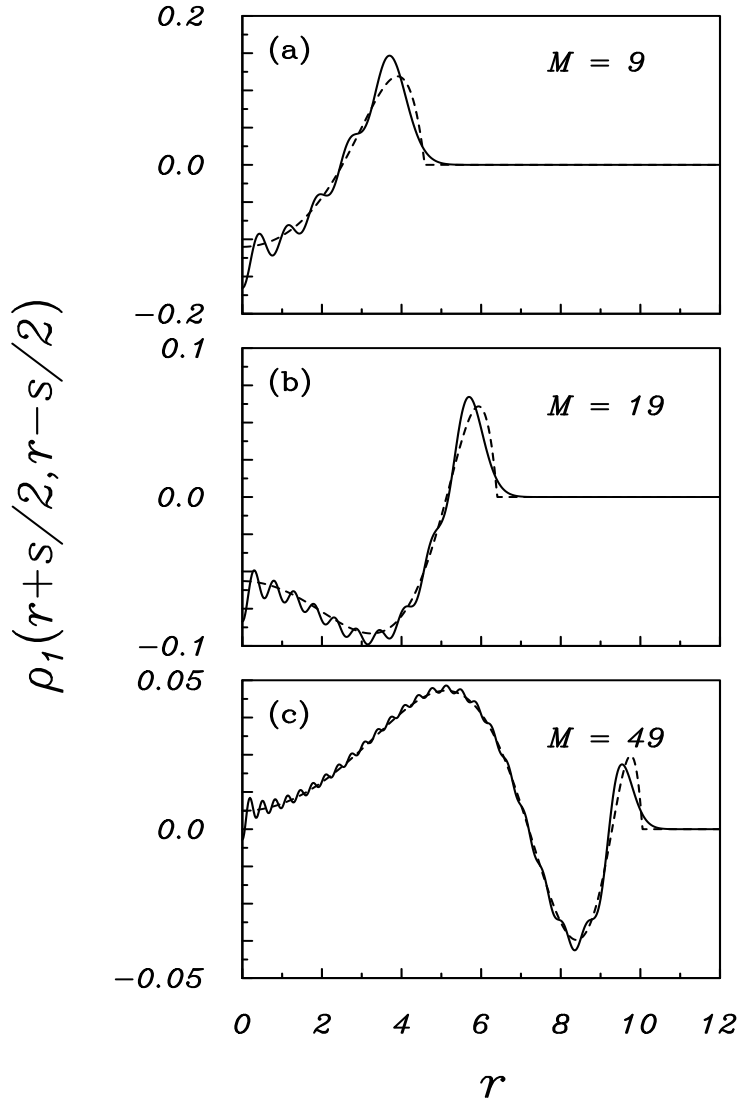}} 
\caption{Comparison of the exact zero temperature density matrix
[Eq.~(20) (solid curves)] with the TFA [Eq.~(31) (dashed curves)] 
for various particle numbers.  Panels (a)-(c) correspond to fixing the
relative coordinate to $\bfs=1$. In all cases, $\rho_1$
is normalized to the central density in the trap, $\rho(0)$.  Note that
at the classical turning point, the TF expression drops abruptly to zero.}
\label{fig2}
\end{figure}
The inverse Laplace transform given by Eq.~(2) (with Eq.~(30) for the Bloch
density matrix) is readily performed, and we find for the 2D harmonically
confined gas 
\bea
\rho_1^{TF}(\bfr_1,\bfr_2) &=& \frac{1}{\pi}
\sqrt{2\left(E_F - V\left(\frac{\bfr_1+\bfr_2}{2}\right)\right)}~
J_1\left[\sqrt{2\left(E_F - V\left(\frac{\bfr_1+\bfr_2}{2}\right)\right)}
|\bfr_1-\bfr_2|\right]\frac{1}{|\bfr_1-\bfr_2|}\nonumber \\
 &=& \frac{1}{\pi}
\sqrt{2\left(E_F - V\left(\bfr\right)\right)}
J_1\left[\sqrt{2(E_F-V(\bfr))}~s\right]\frac{1}{s}~,
\eea
where $J_1(x)$ is a cylindrical Bessel function and 
it is understood that the right-hand side of Eq.~(31) is multiplied
by the unit step function $\Theta(E_F - V(\bfr))$.  Note that Eq.~(31) is also
an exact solution of Eq.~(25).
We immediately observe that for
$E_F \gg V(\bfr)$ (i.e., far from the classical turning point, $r_{cl}$), 
Eq.~(31) becomes effectively identical to the uniform gas result, viz.,
$V(\bfr) = 0$.  Thus, as $N \rightarrow \infty$, $r_{cl} \rightarrow \infty$,
and the first-order density matrix
is essentially unmodified by the trapping potential. 

Now, it is far from obvious by simply looking at the 
functional forms of Eqs.~(20) and (31) that 
the TF and exact expressions are even qualitatively similar. 
Indeed, it is well-known that for
{\em local} properties, such as $\rho(\bfr)$, the TF
and exact densities have very different qualitative behaviour,
especially in low-dimensions and small particle numbers. In particular,
the shell oscillations observed in the exact expressions are not 
reproduced by the (smooth) TF profiles~\cite{brack2}.  
To address this issue, we show in Fig.~1 the exact zero temperature 
density matrix (solid curves)
along with Eq.~(31) (dashed curves) for various particle numbers.  Let us first focus on 
the left panels [i.e., Fig.~1(a)-(c)] which correspond to fixing the center-of-mass
coordinate to $\bfr = 0$.  We observe that for low-particle numbers,
there is a sizable quantitative difference between the two expressions, but by 
$N \sim O(10^3)$, the exact and TF expressions agree
quite well.  In particular, the spatial oscillations in the exact and TFA expressions are
in almost quantitative agreement by Fig.~1(c).  For the right panels 
(which have $\bfr = 3$),
we note the same trend, but now with the agreement between the two 
expressions noticeably improved (i.e., for the same number of particles).  Similar behaviour is
seen for other values of $\bfr$, clearly illustrating that
even at moderate particle numbers, the LDA
is a good description of $\rho_1(\bfr_1,\bfr_2)$, as a function of the relative coordinate $s$.

In Fig.~2(a)-(c) we also compare the exact and TFA of $\rho_1$, but now with the 
{\em relative coordinate}
fixed to $s=1$ and the center-of-mass coordinate allowed to vary.  Note that fixing $s=0$
yields the zero temperature single-particle density, which has already been investigated in
Ref.~\cite{brack2}.  While the overall
spatial behaviour of the two expressions are in agreement, the TFA clearly does not reproduce
the fine spatial oscillations in the exact expression, 
which are associated with shell-filling effects.  
The reason for this, of course, is
due to the fact that the LDA fails to take into account the discrete nature
of the energy level structure of the trap.  We will come back to this
point in the Sec.~IV C below.
Nevertheless, by $M=49$, the agreement between
the exact and TF expression is quite good, except near the classical turning point, where
the TF curves drop abruptly to zero.
%The key point to be taken from this analysis is that quantities derived from
%$\rho_1(\bfr+\bfs/2,\bfr-\bfs/2)$, which depend only on $r$, will not be in quantitative
%agreement with the exact result (at finite $N$).  The reason for this is, of course, 
%due to the fact that the LDA fails to take into account the discrete nature
%of the energy level structure of the trap.  We will come back to this
%important observation in the Sec.~IV C below.

While it is possible to show analytically the reduction of the zero temperature
particle and kinetic energy densities to their TF forms, e.g., $\lim_{N \rightarrow \infty}
\rho(\bfr)^{\rm exact} \rightarrow \rho_{\rm TF}(\bfr)$~\cite{brack1,murthy}, we have
not yet been able to analytically establish a similar result for $\rho(\bfr_1,\bfr_2)$.
%Nevertheless, there is adequate precedent~\cite{brack1,murthy} to suggest that
%is possible, especially in view of the
%its simple analytical form.  
The demonstration of
this result is an interesting problem in its own right.
%%%%%%%%%%%%%%%%%%%%%%%%%%%%%%%%%%%%%%%%%%%%%%%%%%%%%%%%%%%%%%%%%% 
%%%%%%%%%%%%%%%%%%%%%%%%%%%%%%%%%%%%%%%%%%%%%%%%%%%%%%%%%%%%%%%%%%
\section{Exchange energy for a 2D quantum dot: Hartree-Fock approximation}
As a simple application of our results, we now consider the evaluation
of an exact closed form expression for
the HF exchange energy suitable for the study of 2D parabolically confined 
quantum dots.
Here, our motivation for focusing to strictly two dimensions is grounded in
our previous findings for the {\em local} properties of the zero temperature
trapped 2D Fermi gas.  Specifically, in Ref.~\cite{brack2}, we showed
analytically the surprising result that the 2D TF functional 
for the kinetic-energy density (i.e., without gradient corrections), 
when integrated over all space, leads
to the {\em exact} quantum mechanical kinetic energy.  Moreover, we also
demonstrated numerically that if the exact single-particle density is 
inserted into the 2D TF kinetic-energy functional, even
the local shell oscillations are reproduced remarkably well \cite{footnote}.
Thus, the unique local and global properties of the trapped 2D system 
are reason enough for us to focus on two-dimensions.
Nevertheless, we wish to re-emphasize that extending the following
calculations to other 
dimensions requires nothing more than introducing the $d$-dimensional
measure, which is given by
\be
\int d^dr(...) = \frac{\pi^{d/2}}{\Gamma(d/2)}\int_0^\infty x^{d/2-1}dx(...)~.
\ee
For orientation, we will first consider the case for $T=0$ and then generalize 
the result to finite temperatures.   After presenting our analytical expressions,
we will close this section with some illustrative 
numerical results at zero temperature.
%%%%%%%%%%%%%%%%%%%%%%%%%%%%%%%%%%%%%%%%%%%%%%%%%%%%%%%%%%%%%%%%%%%%%%%%%%%%
\subsection{Zero temperature}
Before proceeding with the zero temperature calculation, 
let us first consider a particular class of 
integrals that invariably arise during our manipulations of
$\rho_1(\bfr_1,\bfr_2;T)$, irrespective of dimensionality and temperature, viz.,
\bea
I_{m,n}(\alpha,\beta,\gamma) &=& \int_0^\infty x^\alpha e^{-x}~
L_m^\beta(x) L_n^\gamma(x)~ dx,
%\nonumber \\
%&=& \frac{\Gamma(1+\alpha)\Gamma(n+\gamma+1)\Gamma(\beta-\alpha+m)}
%{\Gamma(m+1)\Gamma(n+1)\Gamma(1+\gamma)\Gamma(\beta-\alpha)}~
%_3F_2[1+\alpha-\beta,-n,1+\alpha;1+\gamma,1+\alpha-\beta-m;1]~,
\eea
where the associated Laguerre polynomials are defined by
\be
L_m^\beta(x) = \sum_{k=0}^{m}\frac{(-1)^k}{k!}\frac{(m+\beta)!}
{(m-k)!(\beta+k)!}x^k~.
\ee
Inserting Eq.~(34) into (33) and integrating term by term, we are left with
a double sum, which can be resummed in closed form to give~\cite{note2}
\be
I_{m,n}(\alpha,\beta,\gamma) =
\frac{\Gamma(1+\alpha)\Gamma(n+\gamma+1)\Gamma(\beta-\alpha+m)}
{\Gamma(m+1)\Gamma(n+1)\Gamma(1+\gamma)\Gamma(\beta-\alpha)}~
_3F_2[1+\alpha-\beta,-n,1+\alpha;1+\gamma,1+\alpha-\beta-m;1]~,
\ee
where $_3F_2[a,b,c;d,e;z]$ is the generalized hypergeometric function~\cite{gr}.
We are now ready to proceed with the calculation of HF exchange energy for 
the 2D quantum dot.

The HF
exchange energy, in the terms of the variables $\bfr$ and $\bfs$, reads
(hereby we set $e = 1$): 
\bea
E_{\rm ex} &=& -\frac{1}{4}\int \int 
\frac{|\rho_1\left(\bfr+\frac{\bfs}{2},\bfr - \frac{\bfs}{2}\right)|^2}{s}
d\bfs d\bfr
\nonumber \\
&=& -\frac{\pi}{2}\int\int_0^\infty 
\left|\rho_1\left(\bfr+\frac{\bfs}{2},\bfr - \frac{\bfs}{2}\right)\right|^2
ds d\bfr~,
\eea
where, from Eq.~(20),
\be 
\left|\rho_1\left(\bfr+\frac{\bfs}{2},\bfr - \frac{\bfs}{2}\right)\right|^2 = 
\frac{4}{\pi^2}\sum_{n=0}^M\sum_{k=0}^M (-1)^{n+k}
L_n(2r^2)L_k(2r^2)e^{-2r^2}L_{M-n}^1(s^2/2)L_{M-k}^1(s^2/2)e^{-s^2/2}~.
\ee
Equation (36) can now be written as
\bea
E_{\rm ex} = -\frac{2}{\pi}\sum_{n=0}^M\sum_{k=0}^M (-1)^{n+k}
\int L_n(2r^2)L_k(2r^2)e^{-2r^2}d\bfr \int_0^\infty  
L_{M-n}^1(s^2/2)L_{M-k}^1(s^2/2)e^{-s^2/2} ds~.
\eea
Going over to the variables $x = s^2/2$, $y=2r^2$ and making use 
of Eq.~(35), viz.,
\bea
I_{m,n}(-1/2,1,1) 
&=& \int_0^\infty x^{-1/2} L_m^1(x) L_n^1(x)e^{-x} dx\nonumber \\
&=& 2(m+1)(n+1)\frac{\Gamma(n+3/2)}{\Gamma(n+2)}~
_3F_2\left[-\frac{1}{2},-m,\frac{1}{2};2,-n-\frac{1}{2};1\right]~,
\eea
we obtain
\bea
E_{\rm ex} &=& -\frac{\sqrt{2}}{\pi}\sum_{n=0}^M\sum_{k=0}^M (-1)^{n+k}
\int L_n(2r^2)L_k(2r^2)e^{-2r^2} d\bfr \int_0^\infty e^{-x} x^{-1/2} 
L_{M-n}^1(x)L_{M-k}^1(x) dx\nonumber \\
&=&
-{\sqrt{2}} \sum_{n=0}^M\sum_{k=0}^M (-1)^{n+k}
(M-n+1)(M-k+1) \frac{\Gamma(M-n+3/2)}{\Gamma(M-n+2)}~
_3F_2\left[-\frac{1}{2},n-M,\frac{1}{2};2,k-M-\frac{1}{2};1\right]\nonumber
\\
&\times&
\int_0^\infty L_n(y)L_k(y)e^{-y} dy\nonumber \\
&=& -\sqrt{2} \sum_{n=0}^{M}(M-n+1)^2\frac{\Gamma(M-n+3/2)}{\Gamma(M-n+2)}~
_3F_2\left[-\frac{1}{2},n-M,\frac{1}{2};2,n-M-\frac{1}{2};1\right]~.
\eea
Equation (40) can be written in the more suggestive form
\be
E_{\rm ex} = \int\varepsilon_{\rm x}(\bfr) d\bfr~,
\ee
whereby we identify the exchange energy density as
\bea
\varepsilon_{\rm x}(\bfr) &=&
-2\frac{\sqrt{2}}{\pi}\sum_{n=0}^M\sum_{k=0}^M(-1)^{n+k}(M-n+1)(M-k+1)
L_n(2r^2)L_k(2r^2)e^{-2r^2}\nonumber \\
&\times& \frac{\Gamma(M-n+\frac{3}{2})}{\Gamma(M-n+2)}~
_3F_2\left[-\frac{1}{2},n-M,\frac{1}{2};2,k-M-\frac{1}{2};1\right]~.
\eea
Note that the first line in Eq.~(42) is $\propto [\rho(\bfr)]^2$.   However,
the terms in the second line of (42)
(i.e., the hypergeometric and Gamma functions) prevent us from writing 
$\varepsilon_{\rm x}(\bfr)$ as a simple functional of
$[\rho(\bfr)]^2$.
%%%%%%%%%%%%%%%%%%%%%%%%%%%%%%%%%%%%%%%%%%%%%%%%%%%%%%%%%%%%%%%%%%%%%%%
\subsection{Finite temperature}
The finite temperature exchange is readily calculated from Eq.~(36)
by using Eq.~(16) for the finite temperature first-order density matrix.
The calculation is entirely analogous to the zero temperature result, with
the central difference being that we now have to evaluate the integral
[see Eq.~(17)]
\bea
I_{m,n}(-1/2,0,0) &=& \int_0^\infty x^{-1/2} L_m(x)L_{n}(x)e^{-x} dx\nonumber \\
&=&\frac{\Gamma(n+1/2)}{\Gamma(n+1)}~
_3F_2\left[-m,\frac{1}{2},\frac{1}{2};\frac{1}{2}-n,1;1\right]~.
\eea
Our final result for the 2D finite temperature exchange energy is given by
\bea
E_{\rm ex}(T) &=&-\frac{\sqrt{2}}{\pi}\sum_{n=0}^\infty
\sum_{n'=0}^\infty \sum_{k=0}^\infty \sum_{k'=0}^\infty
(-1)^{n+n'}F^{(2)}_{n,k}(\mu)F^{(2)}_{n',k'}(\mu)
\frac{\Gamma(k'+1/2)}{\Gamma(k'+1)}~
_3F_2\left[-k,\frac{1}{2},\frac{1}{2};\frac{1}{2}-k',1;1\right]\nonumber \\
&\times& \int L_n(2r^2)L_{n'}(2r^2)e^{-2r^2} d\bfr\nonumber \\
&=&
-\frac{1}{\sqrt{2}}\sum_{n=0}^\infty \sum_{k=0}^\infty \sum_{k'=0}^\infty
F^{(2)}_{n,k}(\mu)F^{(2)}_{n,k'}(\mu)
\frac{\Gamma(k'+1/2)}{\Gamma(k'+1)}~
_3F_2\left[-k,\frac{1}{2},\frac{1}{2};\frac{1}{2}-k',1;1\right]~.
\eea
The finite temperature exchange energy density is similarly given by
\bea
\varepsilon_{\rm x}(\bfr;T) &=&-\frac{\sqrt{2}}{\pi}\sum_{n=0}^\infty
\sum_{n'=0}^\infty \sum_{k=0}^\infty \sum_{k'=0}^\infty
(-1)^{n+n'}F^{(2)}_{n,k}(\mu)F^{(2)}_{n',k'}(\mu)
L_n(2r^2)L_{n'}(2r^2)e^{-2r^2}\nonumber \\
&\times&
\frac{\Gamma(k'+1/2)}{\Gamma(k'+1)}~
_3F_2\left[-k,\frac{1}{2},\frac{1}{2};\frac{1}{2}-k',1;1\right]~.
\eea
While Eq.~(45) looks somewhat unwieldy, it turns out that 
the sums can be truncated relatively quickly, so that the temperature
dependence can be numerically studied, should the need arise.
%%%%%%%%%%%%%%%%%%%%%%%%%%%%%%%%%%%%%%%%%%%%%%%%%%%%%%%%%%%%%%%%%%%%%%%%%%%%
\subsection{Numerical results}
\begin{figure}
\resizebox{5in}{6in}{\includegraphics{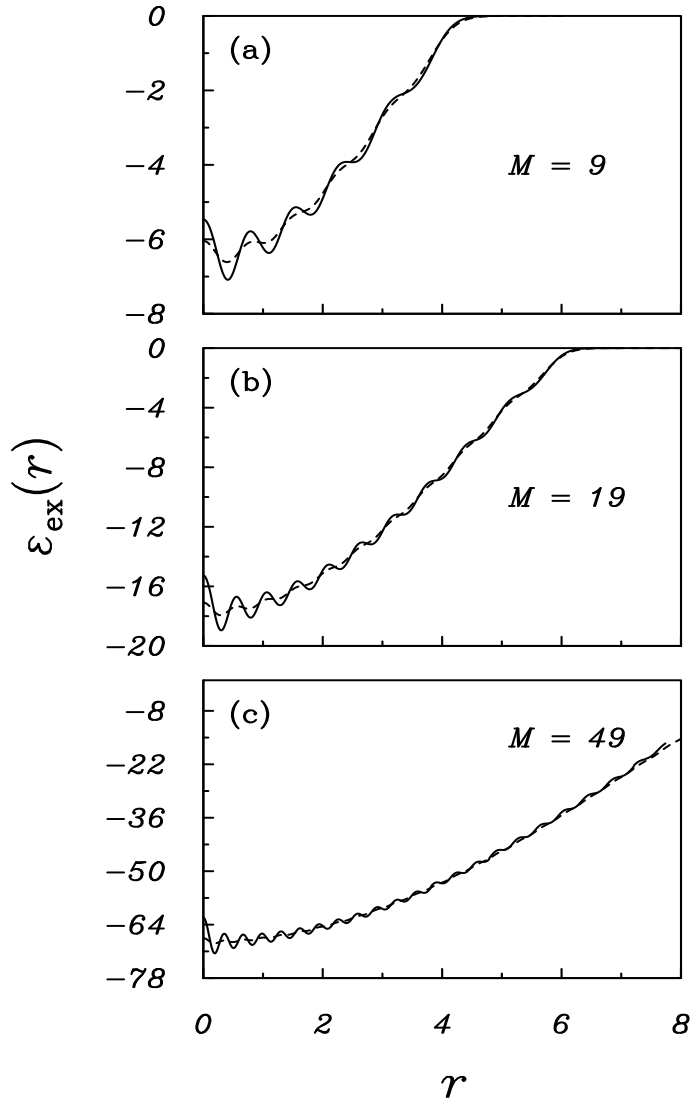}}
\caption{Comparison of the exact [Eq.~(41), solid curves] and
TF [Eq.~(45), dashed curves] 2D exchange energy densities at zero-temperature,
and various particle numbers.  The TF exchange energy density has been generated
by using the {\em exact} single-particle density as input [see Eq.~(46)].}
\label{fig3}
\end{figure}
As mentioned above, the exact zero temperature kinetic energy density, $\tau(\bfr)$, 
for the trapped 2D Fermi gas
agrees remarkably well with the
corresponding TF expression, $T=\pi[\rho(\bfr)]^2/2$, when the
exact $\rho(\bfr)$ is used as input
(see Fig.~3 in Ref.~\cite{brack2}).  It is then
of interest to examine how well the 2D TF exchange energy density compares with
the exact expression, Eq.~(42), in both the small and large $N$-limit.  
Specifically, we will investigate the
applicability of the well-known 2D Dirac exchange energy functional~\cite{note3},
\bea
E^{\rm TF}_{\rm ex} &=& -\frac{4}{3}\sqrt{\frac{2}{\pi}}\int~[\rho(\bfr)]^{3/2}d\bfr\nonumber \\
&=& \int \varepsilon_{\rm ex}^{\rm TF}(\bfr)~d\bfr~,
\eea
for the inhomogeneous 2D gas when the {\em exact} density, 
as given by Eq.~(20) with $\bfr_1 = \bfr_2$, is used as input.
To this end, we present in Fig.~3 (a)-(c) 
the exact (solid curves) and TF (dashed
curves) exchange energy densities for various particle numbers.  
It is clear from this figure that while both expressions have similar qualitative
behaviour, the fine spatial oscillations (i.e., shell-filling effects)
are not well reproduced by the TFA.
This result is entirely expected considering our discussion in Sec.~III in connection
to Fig.~2(a)-(c).
Specifically, in obtaining $\varepsilon_{\rm ex}^{\rm TF}(r)$, we have integrated
over $s$, so that we are left only with the center-of-mass
coordinate $r$.  Therefore, even though the TFA for $\rho_1$ is quite good for the $s$
integration, its failings become apparent upon an examination of the 
exchange density as a function
of $r$; the TFA for $\varepsilon_{\rm ex}(r)$ leads to a {\em smooth}, monotonically
increasing profile because the resulting $r$-dependence is 
$\propto -[\rho^{\rm TF}(\bfr)]^{3/2} =- [(E_F-V(\bfr))/\pi]^{3/2}$ [see Eq.~(46) above].
However, in Fig.~3, 
$\varepsilon_{\rm ex}^{\rm TF}(r)$ has been obtained by using the
exact single-particle density as input, which {\em does} have the 
shell effects encoded in its spatial dependence.   Thus, although shell oscillations can be
included in the TFA of  $\varepsilon^{\rm TF}_{\rm ex}(r)$ (i.e., by using the exact density), 
it is important to note that for a given $r$, 
the weighting given to the
Laguerre polynomials in $[\rho^{\rm exact}(r)]^{3/2}$ {\em is not} the same as the weight
assigned to the Laguerre polynomials in Eq.~(42). 
Consequently, the exchange energy 
densities do not agree as we vary the center-of-mass 
coordinate.  We note in particular that the shell-effects in TFA
are less pronounced than the exact result, and by $M=49$ and $r \gtrsim 3$, are
essentially washed out [see Fig.~3(c)].
It is also worth recounting here, that
by $M=19$ (see Fig.~3 in Ref.~\cite{brack2}),
the deviations between the exact and TF 2D kinetic energy 
densities are essentially nonexistent; this is clearly not the case for
the exchange energy density.   The reasons behind the success of the TFA for the
kinetic energy density are thoughoroughly discussed in Ref.~\cite{murthy}.
Nevertheless, the numerics clearly indicate here that 
as $N\rightarrow \infty$, $\varepsilon_{\rm ex}^{\rm exact}(r) \rightarrow
\varepsilon_{\rm ex}^{\rm TF}(r)$.

While the local behaviour of the exact and TF exchange energy densities
are not well reproduced (as opposed to the comparatively superb agreement
between the exact and TF kinetic energy densities), the exchange energy itself
agrees remarkable well.  To illustrate this, we show in
Table I the exact and TF exchange energies for various numbers of filled shells.
The largest relative percentage error, which occurs at $M+1 = 10$, is only 
$\Delta E/E \simeq 0.5 \%$.
Thus, when {\em global} quantities are considered,
the LDA is an excellent approximation for the inhomogeneous 2D exchange energy,
even for $N \sim O(10^2)$.

     \begin{table}[ht]
     \begin{center} {
     \begin{tabular}{|c|c|c|c|}
     \hline
      $M+1~~$ & $~~E_{\rm ex}^{\rm exact}~~$ & $E_{\rm ex}^{\rm TF}~~$  
      &$\Delta E/E$ \\
       \hline
      10 &~~ -171.71~~~ &~~ -170.81~~~&~~0.5$\%$~~~ \\
      20 &~~ -914.05~~~ &~~ -912.43~~~ &~~0.2 $\%$~~~\\
      50 &~~~~ -8703.06~~~ & ~~~~-8699.51~~~&~~~~0.04 $\%$~~~ \\
      \hline
     \end{tabular} }
     \end{center}
     \caption{Comparison of the zero temperature 
     exchange energy for the exact [Eq.~(40)] and TF
      [Eq.~(46)] expressions for various numbers of filled shells.
      The last column displays the relative percentage 
      error in the two quantities.}
     \label{turns}
     \end{table}
%%%%%%%%%%%%%%%%%%%%%%%%%%%%%%%%%%%%%%%%%%%%%%%%%%%%%%%%%%%%%%%%%%%%%%%%%%%%%%
%%%%%%%%%%%%%%%%%%%%%%%%%%%%%%%%%%%%%%%%%%%%%%%%%%%%%%%%%%%%%%%%%%%%%%%%%%%%%%
\section{Summary and future work}
The main result of this paper can be summarized by Eq.~(22), which defines the exact,
$d$-dimensional, finite temperature first-order density matrix of an ideal gas of 
harmonically confined fermions.  This expression should prove
to be of interest in the general area of the DFT of inhomogeneous Fermi systems
at both zero and finite temperatures.  In this paper, we have used it to illustrate that (in 2D)
the LDA is an excellent approximation for the 
the off-diagonal first-order density matrix in the large-$N$ limit. 
We have also obtained a simple, closed form expression for the finite temperature
exchange energy density for a 2D parabolically confined quantum dot.  In the spirit of our
previous work~\cite{brack2}, we have utilized this exact expression to test the
validity of the LDA for the exchange energy density.  In contrast to our earlier 
findings~\cite{brack2}, 
the 2D TF exchange energy functional {\em does not} reproduce the shell effects  of the
exact result very well.  Nevertheless, when the TF exchange
energy density is integrated over all space, we find that the 
resulting exchange energy is always within $\lesssim 0.5 \%  $ of the exact result.

We wish to point out that the
utility of our results are not limited to the topics discussed
in this paper.   The simple, analytical expressions that we have provided will be
very useful in the obtaining other closed form expressions of interest to both theorists
working in formal DFT,
and experimentalists studying e.g., ultra-cold trapped fermions.  
One example that comes to mind is the 2D weakly interacting trapped Fermi gas.  In
Ref.~\cite{brack1}, the use of a contact psuedopotential (in 2D) led to the surprising
discovery that there is {\em no splitting} between states with different angular
momentum values $l$ in a given shell at $T=0$.  This occurs in spite of the fact that
the perturbation interaction does not preserve the $SU(2)$ symmetry of the system.
Whether this result still holds true for a finite-range psuedopotential is an 
interesting question, which can be addressed using our analytical expression for
$\rho_1(\bfr_1,\bfr_2)$.  
%Experimentalists may be interested in exact expressions of other
%quantities such as the momentum density $n({\bf p})$~\cite{howard3}, which is
%experimentally accessible by measuring the lineshape in Compton scattering.
%Having an exact expression, from which experimental data can be compared, could provide
%insight into the role of dimensionality and interactions in these very dilute inhomogeneous 
%Fermi gases.  
In addition, the close connection between trapped Fermi gases and the theory of 
nuclear structure suggests that our findings will be relavent in the 
area of nuclear physics (e.g., in the Fermi gas model of the nucleus).  Finally, our
work here should also be useful in the physics of metal clusters, which represent an
intermediate stage in the transition from small molecules to bulk solids or liquids.

\begin{acknowledgments}
It is a pleasure to thank Drs.~R.K. Bhaduri, M. Brack, and M.V.N Murthy for useful
discussions.  I would like to acknowledge financial support from Dr.~R.K. Bhaduri
through a grant from the National Sciences and the Engineering
Research Council of Canada (NSERC).
\end{acknowledgments}
\bibliographystyle{pra}

\begin{thebibliography}{20}
\bibitem{footnote} Similar agreement is also found for other dimensions.  However, it
is only the 2D system whose TF kinetic energy functional (with the exact density used
as input) integrates to the exact quantum mechanical kinetic energy.
\bibitem{note2} We have used the mathematical software package,  
Maple$^\copyright$, to evaluate this quantity.  We have also checked this result by hand.
\bibitem{note3} This functional can easily be derived by using Eq.~(31) in Eq.~(36), and
identifying $\rho(\bfr) = (E_F-V(\bfr))/\pi$.
\end{thebibliography}

\end{document}